\documentclass[runningheads]{llncs}
\usepackage{graphicx}
\usepackage{listings}
\usepackage{subfig}
\usepackage{amsmath}
\usepackage{wrapfig}
\usepackage{bm}

\usepackage{amsfonts}
\usepackage{stmaryrd}
\usepackage{amssymb}

\usepackage{pifont}
\newcommand{\cmark}{\ding{51}}%
\newcommand{\xmark}{\ding{55}}%

\usepackage{mathtools}
\usepackage{mathpartir}
\usepackage{tabularx}
\usepackage{diagbox}
\newcolumntype{C}{>{\centering\arraybackslash}X}

\newcommand{\C}[1]{\texttt{#1}}
\newcommand{\F}[1]{\mathsf{#1}}
\newcommand{\M}[1]{\mathcal{#1}}

\newcommand{\RULE}[2]{\frac{\begin{array}{c}#1\end{array}}
                           {\begin{array}{c}#2\end{array}}}

\begin{document}
\title{Automated Parameterized Verification of CRDTs (Extended Version)}
%
%
\author{Kartik Nagar \and
Suresh Jagannathan}
%
\institute{Purdue University, USA\\
\email{\{nagark,suresh\}@cs.purdue.edu}} 

%
\maketitle              
\begin{abstract}
  
Maintaining multiple replicas of data is crucial to achieving
scalability, availability and low latency in distributed
applications. \emph{Conflict-free Replicated Data Types} (CRDTs) are
important building blocks in this domain because they are designed to
operate correctly under the myriad behaviors possible in a
weakly-consistent distributed setting.  Because of the possibility of
concurrent updates to the same object at different replicas, and the
absence of any ordering guarantees on these updates,
\emph{convergence} is an important correctness criterion for CRDTs.
This property asserts that two replicas which receive the same set of
updates (in any order) must nonetheless converge to the same state.
One way to prove that operations on a CRDT converge is to show that
they commute since commutative actions by definition behave the same
regardless of the order in which they execute.  In this paper, we
present a framework for automatically verifying convergence of CRDTs under different
weak-consistency policies.  Surprisingly, depending upon the
consistency policy supported by the underlying system, we show that
not all operations of a CRDT need to commute to achieve convergence.
We develop a proof rule parameterized by a consistency specification
based on the concepts of \emph{commutativity modulo consistency
  policy} and \emph{non-interference to commutativity}.  We describe
the design and implementation of a verification engine equipped with
this rule and show how it can be used to provide the first automated
convergence proofs for a number of challenging CRDTs, including sets,
lists, and graphs.

\end{abstract}

\section{Introduction}

For distributed applications, keeping a single copy of data at one
location or multiple fully-synchronized copies (i.e. state-machine
replication) at different locations, makes the application susceptible
to loss of availability due to network and machine failures. On the
other hand, having multiple un-synchronized replicas of the data
results in high availability, fault tolerance and uniform low latency,
albeit at the expense of consistency. In the latter case, an update
issued at one replica can be asynchronously transmitted to other
replicas, allowing the system to operate continuously even in the
presence of network or node failures~\cite{GI02}. However, mechanisms
must now be provided to ensure replicas are kept consistent with each
other in the face of concurrent updates and arbitrary re-ordering of
such updates by the underlying network.

Over the last few years, \emph{Conflict-free Replicated Datatypes}
(CRDTs) \cite{SH11a,SH11b,PR18} have emerged as a popular solution to
this problem. In op-based CRDTs, when an operation on a CRDT instance
is issued at a replica, an \emph{effector} (basically an update
function) is generated locally, which is then asynchronously
transmitted (and applied) at all other replicas.\footnote{In this
  work, we focus on the op-based CRDT model; however, our technique
  naturally extends to state-based CRDTs since they can be
  emulated by an op-based model\cite{SH11a}} Over the years, a number of CRDTs
have been developed for common datatypes such as maps, sets, lists,
graphs, etc.

The primary correctness criterion for a CRDT implementation is
\emph{convergence} (sometimes called \emph{strong eventual
  consistency}~\cite{GO17,SH11a} ({\sf SEC})): two replicas which have
received the same set of effectors must converge to the same CRDT
state.  Because of the weak default guarantees assumed to be provided
by the underlying network, however, we must consider the possibility
that effectors can be applied in arbitrary order on different
replicas, complicating correctness arguments.  This complexity is
further exacerbated because CRDTs impose no limitations on how often
they are invoked, and may assume additional properties on network
behaviour~\cite{LL11} that must be taken into account when formulating
correctness arguments.

Given these complexities, verifying convergence of operations in a
replicated setting has proven to be challenging and
error-prone~\cite{GO17}.  In response, several recent efforts have
used mechanized proof assistants to yield formal machine-checked
proofs of correctness~\cite{GO17,ZE14}.  While mechanization clearly
offers stronger assurance guarantees than handwritten proofs, it still
demands substantial manual proof engineering effort to be successful.
In particular, correctness arguments are typically given in terms of
constraints on CRDT states that must be satisfied by the underlying
network model responsible for delivering updates performed by other
replicas.  Relating the state of a CRDT at one replica with the
visibility properties allowed by the underlying network has typically
involved constructing an intricate simulation argument or crafting a
suitably precise invariant to establish convergence.  This level of
sophisticated reasoning is required for every CRDT and consistency
model under consideration.  There is a notable lack of techniques
capable of reasoning about CRDT correctness under different weak
consistency policies, even though such techniques exist for other
correctness criteria such as preservation of state invariants
\cite{GO16,HO19} or serializability \cite{BG16,NJ18} under weak
consistency.

To overcome these challenges, we propose a novel \emph{automated}
verification strategy that does not require complex proof-engineering
of handcrafted simulation arguments or invariants.  Instead, our
methodology allows us to directly connect constraints on events
imposed by the consistency model with constraints on states required
to prove convergence.  Consistency model constraints are extracted
from an axiomatization of network behavior, while state constraints
are generated using reasoning principles that determine the
\emph{commutativity} and \emph{non-interference} of sequences of
effectors, subject to these consistency constraints.  Both sets of
constraints can be solved using off-the-shelf theorem provers.
Because an important advantage of our approach is that it is
parametric on weak consistency schemes, we are able to analyze the
problem of CRDT convergence under widely different consistency
policies (e.g., eventual consistency, causal consistency, parallel
snapshot isolation ({\sf PSI})~\cite{SO11}, among others), and for the
first time verify CRDT convergence under such stronger models
(efficient implementations of which are supported by real-world data
stores).  A further pleasant by-product of our approach is a pathway
to take advantage of such stronger models to simplify existing CRDT
designs and allow composition of CRDTs to yield new instantiations for
more complex datatypes.

The paper makes the following contributions:
\begin{enumerate}

\item We present a proof methodology for verifying the correctness of
  CRDTs amenable to automated reasoning.

\item We allow the proof strategy to be parameterized on a weak
  consistency specification that allows us to state correctness
  arguments for a CRDT based on constraints imposed by these
  specifications.

\item We experimentally demonstrate the effectiveness of
our proposed verification strategy on a
  number of challenging CRDT implementations across multiple
  consistency schemes.

\end{enumerate}

\noindent Collectively, these contributions yield (to the best of our
knowledge) the first automated and parameterized proof methodology for
CRDT verification.

The remainder of the paper is organized as follows.  In the next
section, we provide further motivation and intuition for our approach.
Sec.~\ref{sec:opsem} formalizes the problem definition, providing an
operational semantics and axiomatizations of well-known consistency
specifications.  Sec.~\ref{sec:proofrule} describes our proof strategy
for determining CRDT convergence that is amenable to automated
verification.  Sec.~\ref{sec:impl} provides details about our
implementation and experimental results justifying the effectiveness
of our framework.  Secs.~\ref{sec:related} presents
related work and conclusions.




\section{Illustrative Example}

\begin{wrapfigure}{l}{.42\textwidth}
  \small
  \vspace*{-.35in}
\begin{lstlisting}
  S$\in \bbbp(E)$
Add(a):S  $\lambda$S'.S'$\cup${a}
Remove(a):S  $\lambda$S'.S'$\setminus${a}
Lookup(a):S  a $\in$ S
\end{lstlisting}
\caption{\small A simple {\sf Set} CRDT definition.}
\label{fig:ex(a)}
\vspace*{-.25in}
\end{wrapfigure}

We illustrate our approach using a {\sf Set} CRDT specification as a
running example. A CRDT $(\Sigma,O,\sigma_{\F{init}})$ is
characterized by a set of states $\Sigma$, a set of operations $O$ and
an initial state $\sigma_{\F{init}} \in \Sigma$, where each operation
$o \in O$ is a function with signature $\Sigma \rightarrow (\Sigma
\rightarrow \Sigma)$. The state of a CRDT is replicated, and when
operation $o$ is issued at a replica with state $\sigma$, the effector
$o(\sigma)$ is generated, which is immediately applied at the local
replica (which we also call the \emph{source} replica) and transmitted
to all other replicas, where it is subsequently applied upon receipt.

Additional constraints on the order in which effectors can be received and
applied at different replicas are specified by a consistency policy,
discussed below.  In the absence of any such additional constraints,
however, we assume the underlying network only offers \emph{eventually
  consistent} guarantees - all replicas eventually receive all
effectors generated by all other replicas, with no constraints  on the
order in which these effectors are received.

Consider the simple {\sf Set} CRDT definition shown in
Fig.~\ref{fig:ex(a)}.  Let $E$ be an arbitrary set of elements. The
state space $\Sigma$ is $\bbbp(E)$. \texttt{Add(a):S} denotes the
operation \texttt{Add(a)} applied on a replica with state \C{S}, which
generates an effector which simply adds \C{a} to the state of all
other replicas it is applied to.  Similarly, \C{Remove(a):S} generates
an effector that removes \C{a} on all replicas to which it is applied.
\C{Lookup(a):S} is a query operation which checks whether the queried
element is present in the source replica \C{S}.

A CRDT is \emph{convergent} if during any execution, any two replicas
which have received the same set of effectors have the same state.
Our strategy to prove convergence is to show that any two effectors of
the CRDT pairwise commute with each other modulo a consistency policy,
i.e. for two effectors $e_1$ and $e_2$, $e_1 \circ e_2 = e_2 \circ
e_1$. Our simple {\sf Set} CRDT clearly does not converge when
executed on an eventually consistent data store since the effectors
$e_1 =$ \C{Add(a):S}$_1$ and $e_2 =$ \C{Remove(a):S}$_2$ do not
commute, and the semantics of eventual consistency imposes no
additional constraints on the visibility or ordering of these
operations that could be used to guarantee convergence.  For example, if
$e_1$ is applied to the state at some replica followed by the
application of $e_2$, the resulting state does not include the element
\C{a}; conversely, applying $e_2$ to a state at some replica followed
by $e_1$ leads to a state that does contain the element \C{a}.
\begin{wrapfigure}{r}{0.4\textwidth}
  \small
  \vspace*{-.3in}
\begin{lstlisting}
  S$\in \bbbp(E \times I)$
Add(a,i):S
  $\lambda$S'.S'$\cup${(a,i)}
Remove(a):S 
  $\lambda$S'.S'$\setminus${(a,i):(a,i)$\in$S}
Lookup(a):S
  $\exists$(a,i)$\in$A
\end{lstlisting}
\caption{\small A definition of an {\sf ORSet} CRDT.}\label{fig:ex(b)}
\vspace*{-.3in}
\end{wrapfigure}
However, while commutativity is a sufficient property to show
convergence, it is not always a necessary one.  In particular,
different consistency models impose different constraints on the
visibility and ordering of effectors that can obviate the need to
reason about their commutativity. For example, if the consistency
model enforces \C{Add(a)} and \C{Remove(a)} effectors to be 
applied in the same order at all replicas, then the {\sf Set} CRDT 
will converge. As we will demonstrate later, the PSI consistency
model exactly matches this requirement.
   To further illustrate this, consider the definition of the {\sf
  ORSet} CRDT shown in Fig.~\ref{fig:ex(b)}.  Here, every element is
tagged with a unique identifier (coming from the set
$I$). \C{Add(a,i):S} simply adds the element \C{a} tagged with
\C{i}\footnote{Assume that every call to \C{Add} uses a unique
  identifier, which can be easily arranged, for example by keeping a
  local counter at every replica which is incremented at every
  operation invocation, and using the id of the replica and the value
  of the counter as a unique identifier}, while \C{Remove(a):S}
returns an effector that when applied to a replica state will remove
all tagged versions of \C{a} that were present in \C{S}, the source
replica.

Suppose $e_1=$\C{Add(a,i):S}$_1$ and $e_2=$\C{Remove(a):S}$_2$.  If it
is the case that \C{S}$_2$ does not contain \C{(a,i)}, then these two
effectors are guaranteed to commute because $e_2$ is unaware of
\C{(a,i)} and thus behaves as a no-op with respect to effector $e_1$
when it is applied to any replica state.  Suppose, however, that
$e_1$'s effect was visible to $e_2$; in other words, $e_1$ is applied
to \C{S}$_2$ before $e_2$ is generated.  There are two possible scenarios
that must be considered.  (1) Another replica (call it \C{S'}) has $e_2$ applied
before $e_1$.  Its final state reflects the effect of the \C{Add}
operation, while \C{S}$_2$'s final state reflects the effect of applying the
\C{Remove}; clearly, convergence is violated in this case.  (2) All
replicas apply $e_1$ and $e_2$ in the same order; the interesting case here is
when the effect
of $e_1$ is always applied before $e_2$ on every replica.  The constraint
that induces an effector order between $e_1$ and $e_2$ on every
replica as a consequence of $e_1$'s visibility to $e_2$ on \C{S}$_2$
is supported by a causally consistent distributed storage model.  Under
causal consistency, whenever $e_2$ is applied to a replica state, we
are guaranteed that $e_1$'s effect, which adds \C{(a,i)} to the state,
would have occurred.  Thus, even though $e_1$
and
\begin{wrapfigure}{l}{0.53\textwidth} \small
\vspace*{-.3in}
\begin{lstlisting}
  S$\in \bbbp(E \times I) \times \bbbp(E \times I)$
Add(a,i):(A,R)
  $\lambda$(A',R').(A'$\cup${(a,i)},R')
	
Remove(a):(A,R) 
  $\lambda$(A',R').(A',R'$\cup${(a,i):(a,i)$\in$A}
  
Lookup(a):(A,R)
  $\exists$(a,i)$\in$A$\wedge$(a,i)$\notin$R
\end{lstlisting}
\caption{\small A variant of the {\sf ORSet} using tombstones.}\label{fig:ex(c)}
\vspace*{-.3in}
\end{wrapfigure}
$e_2$ do not commute when applied to an arbitrary state, their
execution under causal consistency nonetheless allows us to show that
all replica states converge.  The essence of our proof methodology is
therefore to reason about \emph{commutativity modulo consistency} - it
is only for those CRDT operations unaffected by the constraints
imposed by the consistency model that proving commutativity is
required.  Consistency properties that affect the visibility of
effectors are instead used to guide and simplify our analysis.
Applying this notion to pairs of effectors in arbitrarily long
executions requires incorporating commutativity properties under a
more general induction principle to allow us to generalize the
commutativity of effectors in bounded executions to the unbounded
case.  This generalization forms the heart of our automated verification
strategy.
Fig.~\ref{fig:ex(c)} defines an {\sf ORSet} with ``tombstone'' markers
used to keep track of deleted elements in a separate set.
Our proof methodology is sufficient to automatically show
that this CRDT converges under {\sf EC}.

\section{Problem Definition}
\label{sec:opsem}

In this section, we formalize the problem of determining convergence
in CRDTs parametric to a weak consistency policy. First, we define a
general operational semantics to describe all valid executions of a
CRDT under any given weak consistency policy. As stated earlier, a
CRDT program $\mathcal{P}$ is specified by the tuple $(\Sigma, O,
\sigma_{\F{init}})$. Here, we find it to convenient to define an
operation $o \in O$ as a function $(\Sigma \times (\Sigma \rightarrow
\Sigma)^*) \rightarrow (\Sigma \rightarrow \Sigma)$. Instead of
directly taking as input a generating state, operations are now
defined to take as input a start state and a sequence of
effectors. The intended semantics is that the sequence of effectors
would be applied to the start state to obtain the generating
state. Using this syntax allows us simplify the presentation of the
proof methodology in the next section, since we can abstract a history of
effectors into an equivalent start state. 

Formally, if $\hat{o}:\Sigma \rightarrow (\Sigma \rightarrow \Sigma)$
was the original op-based definition, then we define the operation 
$o:(\Sigma \times (\Sigma \rightarrow
\Sigma)^*) \rightarrow (\Sigma \rightarrow \Sigma)$ as follows:
\begin{mathpar}
\begin{array}{rrcl}
\forall \sigma. &\ o(\sigma, \epsilon) & = & \hat{o}(\sigma)\\
\forall \sigma,\pi,f. &\ o(\sigma, \pi f) & = & o(f(\sigma), \pi)
\end{array}
\end{mathpar}

\noindent Note that $\epsilon$ indicates the empty sequence.
Hence, for all states $\sigma$ and sequence of functions $\pi$, 
we have $o(\sigma, \pi) = \hat{o}(\pi(\sigma))$.

To define the operational semantics, we abstract away from the concept
of replicas, and instead maintain a global pool of effectors. A new
CRDT operation is executed against a CRDT state obtained by first
selecting a subset of effectors from the global pool and then applying
the elements in that set in some non-deterministically chosen
permutation to the initial CRDT state. The choice of effectors and
their permutation must obey the weak consistency policy
specification. Given a CRDT $\mathcal{P} = (\Sigma, O,
\sigma_{\F{init}})$ and a weak consistency policy $\Psi$, we define a
\textbf{labeled transition system} $\mathcal{S}_{\mathcal{P}, \Psi} =
(\mathcal{C},\rightarrow)$, where $\mathcal{C}$ is a set of
configurations and $\rightarrow$ is the transition relation. A
\textbf{configuration} $c = (\Delta,\F{vis},\F{eo})$ consists of three
components : $\Delta$ is a set of events, $\F{vis} \subseteq
\Delta \times \Delta$ is a \emph{visibility} relation, and
$\F{eo}\subseteq \Delta \times \Delta$ is a global \emph{effector
  order} relation (constrained to be anti-symmetric). An
\textbf{event} $\eta \in \Delta$ is a tuple $(\F{eid}, o, \sigma_s,
\Delta_r, \F{eo})$ where $\F{eid}$ is a unique event id, $o \in O$ is
a CRDT operation, $\sigma_s \in \Sigma$ is the start CRDT state,
$\Delta_r$ is the set of events visible to $\eta$ (also called the
history of $\eta$), and $\F{eo}$ is a total order on the events in
$\Delta_r$ (also called the local effector order relation). We assume
projection functions for each component of an event (for example 
$\sigma_s(\eta)$ projects the start state of the event $\eta$).

Given an event $\eta = (\F{eid}, o, \sigma_s, \Delta_r, \F{eo})$, we define
$\eta^{e}$ to be the \textbf{effector} associated with the event. This
effector is obtained by executing the CRDT operation $o$ against the
start CRDT state $\sigma_s$ and the sequence of effectors obtained from the
events in $\Delta_r$ arranged in the reverse order of $\F{eo}$. Formally,
\begin{equation}
\eta^{e} = \begin{cases}
o( \sigma_{s}, \epsilon) & \mbox{if } \Delta_r = \phi\\
o( \sigma_{s}, \prod_{i=1}^{k} \eta_{P(i)}^{e}) & \mbox{if } \Delta_r = \{\eta_1, \ldots, \eta_k\} \mbox{ where } P:\{1,\ldots,k\} \rightarrow \{1,\ldots,k\}\\ & \forall i,j. i < j \Rightarrow (\eta_{P(j)}, \eta_{P(i)}) \in \F{eo}
\end{cases}
\end{equation}

\noindent In the above definition, when $\Delta_r$ is non-empty, we define a
permutation $P$ of the events in $\Delta_r$ such that the permutation
order is the inverse of the effector order $\F{eo}$. This ensures that
if $(\eta_i,\eta_j) \in \F{eo}$, then $\eta_j^e$ occurs before
$\eta_i^e$ in the sequence passed to the CRDT operation $o$,
effectively applying $\eta_i^e$ before $\eta_j^e$ to obtain the
generating state for $o$.

The following rule describes the transitions allowed in $\mathcal{S}_{\mathcal{P}, \Psi}$:
\begin{mathpar}
\footnotesize
\RULE
{\Delta_r \subseteq \Delta \quad o \in O \quad \sigma_s \in \Sigma \quad \F{eo}_r  \text{ is a total order on } \Delta_r \\ \F{eo} \subseteq \F{eo}_r \quad \F{fresh} \textsf{ id} \quad  \eta = (\F{id}, o, \sigma_s, \Delta_r, \F{eo}) \\ \Delta' = \Delta \cup \{\eta\} \quad \F{vis}' = \F{vis} \cup \{(\eta', \eta)\ |\ \eta' \in \Delta_r\} \quad \Psi(\Delta', \F{vis}', \F{eo}') }
{(\Delta, \F{vis}, \F{eo}) \xrightarrow{\eta} (\Delta', \F{vis}', \F{eo}') }
\end{mathpar}

The rule describes the effect of executing a new operation $o$, which
begins by first selecting a subset of already completed events
($\Delta_r$) and a total order $\F{eo}_r$ on these events which obeys
the global effector order $\F{eo}$. This mimics applying the operation
$o$ on an arbitrary replica on which the events of $\Delta_r$ have
been applied in the order $\F{eo}_r$. A new event ($\eta$)
corresponding to the issued operation $o$ is computed, which is used to label
the transition and is also 
added to the current configuration. All the events in $\Delta_r$ are
visible to the new event $\eta$, which is reflected in the new
visibility relation $\F{vis}'$. The system moves to the new
configuration $(\Delta', \F{vis}', \F{eo}')$ which must satisfy the
consistency policy $\Psi$.  Note that even though the general
transition rule allows the event to pick any arbitrary start state $\sigma_s$,
we restrict the start state of all events in a \textbf{well-formed execution}
to be the initial CRDT state $\sigma_{\F{init}}$, i.e. the state in
which all replicas begin their execution. A trace of
$\mathcal{S}_{\mathcal{P}, \Psi}$ is a sequence of transitions. 
Let $\llbracket \mathcal{S}_{\mathcal{P}, \Psi} \rrbracket$ be the set of all finite traces.  Given
a trace $\tau$, $L(\tau)$ denotes all events (i.e. labels) in $\tau$.

\begin{definition}[Well-formed Execution]
A trace $\tau \in \llbracket \mathcal{S}_{\mathcal{P}, \Psi} \rrbracket$ is a well-formed execution if it begins from the empty configuration $C_{\F{init}} = (\{\},\{\},\{\})$ and $\forall \eta \in L(\tau)$, $\sigma_s(\eta) = \sigma_{\F{init}}$.
\end{definition}
%

Let $\M{WF}(\M{S}_{\M{P}, \Psi})$ denote all
well-formed executions of $\M{S}_{\M{P}, \Psi}$. The \textbf{consistency policy} $\Psi(\Delta, \F{vis},
\F{eo})$ is a formula constraining the events in $\Delta$ and 
relations $\F{vis}$ and $\F{eo}$ defined over
these events. Below, we illustrate how to express certain well-known
consistency policies in our framework:

\begin{center}
\small
\begin{tabular}{| l | c |}
\hline
\textbf{Consistency Scheme} & $\bm{\Psi(\Delta, \F{vis}, \F{eo})}$\\
\hline \hline
{\sf Eventual Consistency} \cite{BA13} & $\forall \eta,\eta' \in \Delta. \neg \F{eo}(\eta,\eta')$\\ \hline
{\sf Causal Consistency} \cite{LL11} & $\forall \eta,\eta' \in \Delta. \F{vis}(\eta,\eta') \Leftrightarrow \F{eo}(\eta,\eta') $ \\ 
& $\wedge \forall \eta,\eta', \eta'' \in \Delta. \F{vis}(\eta, \eta') \wedge \F{vis}(\eta', \eta'') \Rightarrow \F{vis}(\eta, \eta'')$ \\ \hline
{\sf RedBlue Consistency} ($O_r$) \cite{LI12} & $\forall \eta,\eta' \in \Delta. o(\eta) \in O_r \wedge o(\eta') \in O_r \wedge \F{vis}(\eta,\eta') \Leftrightarrow \F{eo}(\eta,\eta') $ \\ 
& $\wedge \forall \eta,\eta' \in \Delta. o(\eta) \in O_r \wedge o(\eta') \in O_r \Rightarrow \F{vis}(\eta,\eta') \vee \F{vis}(\eta',\eta)$ \\ \hline
{\sf Parallel Snapshot Isolation} \cite{SO11} & $\forall \eta,\eta' \in \Delta. (\F{Wr}(\eta^e) \cap \F{Wr}(\eta^{'e}) \neq \phi \wedge \F{vis}(\eta,\eta')) \Leftrightarrow \F{eo}(\eta,\eta')$ \\
& $\wedge \forall \eta,\eta' \in \Delta. \F{Wr}(\eta^e) \cap \F{Wr}(\eta^{'e}) \neq \phi \Rightarrow \F{vis}(\eta,\eta') \vee \F{vis}(\eta',\eta)$ \\ \hline
{\sf Strong Consistency} & $\forall \eta,\eta' \in \Delta. \F{vis}(\eta,\eta') \Leftrightarrow \F{eo}(\eta,\eta')$ \\
&  $\wedge \forall \eta,\eta' \in \Delta. \F{vis}(\eta,\eta') \vee \F{vis}(\eta',\eta)$\\
\hline
\end{tabular}
\end{center}

For {\sf Eventual Consistency} ({\sf EC}) \cite{BA13}, we do not place
any constraints on the visibility order and require the global
effector order to be empty.  This reflects the fact that in {\sf EC},
any number of events can occur concurrently at different replicas, and
hence a replica can witness any arbitrary subset of events which may
be applied in any order.  In {\sf Causal Consistency} ({\sf CC})
\cite{LL11}, an event is applied at a replica only if all causally
dependent events have already been applied. An event $\eta_1$ is
causally dependent on $\eta_2$ if $\eta_1$ was generated at a replica
where either $\eta_2$ or any other event causally dependent on
$\eta_2$ had already been applied. The visibility relation $\F{vis}$
captures causal dependency, and by making $\F{vis}$ transitive, we
ensure that all causal dependencies of events in $\Delta_r$ are also
present in $\Delta_r$ (this is because in the transition rule, $\Psi$
is checked on the updated visibility relation which relates events in
$\Delta_r$ with the newly generated event). Further, causally
dependent events must be applied in the same order at all replicas,
which we capture by asserting that $\F{vis}$ implies $\F{eo}$.  In
{\sf RedBlue Consistency} ({\sf RB}) \cite{LI12}, a subset of CRDT
operations ($O_r \subseteq O$) are synchronized, so that they must
occur in the same order at all replicas. We express {\sf RB} in our framework by requiring the
visibility relation to be total among events whose operations are in
$O_r$.  In {\sf Parallel Snapshot
  Isolation} ({\sf PSI}) \cite{SO11}, two events which conflict with
each other (because they write to a common variable) are not allowed
to be executed concurrently, but are synchronized across all replicas
to be executed in the same order. Similar to \cite{GO16}, we assume
that when a CRDT is used under {\sf PSI}, its state space $\Sigma$ is
a map from variables to values, and every operation generates an
effector which simply writes to certain variables.  We assume that
$\F{Wr}(\eta^e)$ returns the set of variables written by the effector
$\eta^e$, and express {\sf PSI} in our framework by requiring that
events which write a common variable are applied in the same order
(determined by their visibility relation) across all replicas;
furthermore, the policy requires that the visibility operation among
such events is total. Finally, in {\sf Strong Consistency}, the
visibility relation is total and all effectors are applied in the same
order at all replicas.

Given an execution $\tau \in
\llbracket \M{S}_{\M{P}, \Psi} \rrbracket$ and a transition $C
\xrightarrow{\eta} C'$ in $\tau$, we associate a set of replica states
$\Sigma_{\eta}$ that the event can potentially witness, by considering
all permutations of the effectors visible to $\eta$ which obey the global
effector order, when applied to
the start state $\sigma_{s}(\eta)$. Formally, this is defined
as follows, assuming $\eta = (\F{eid}, o, \sigma_{s}, \{\eta_1, \ldots, \eta_k\},
\F{eo}_r)$ and $C = (\Delta, \F{vis}, \F{eo})$):
\begin{mathpar}
\begin{array}{lcl}
\Sigma_{\eta} & = & \{\eta_{P(1)}^e \circ \eta_{P(2)}^e \circ \ldots \circ \eta_{P(k)}^e(\sigma_{s})\ |\ P : \{1, \ldots, k\} \rightarrow \{1, \ldots, k\},\\ & & \F{eo}_{P} \text{ is a total order }, i < j \Rightarrow (\eta_{P(j)}, \eta_{P(i)}) \in \F{eo}_{P},\ \F{eo} \subseteq \F{eo}_{P} \}
\end{array}
\end{mathpar} 

\noindent In the above definition, for all valid local effector orders
$\F{eo}_{P}$, we compute the CRDT states obtained on applying those
effectors on the start CRDT state, which constitute
$\Sigma_{\eta}$. The original event $\eta$ presumably would have
witnessed one of these states.

 \begin{definition}[Convergent Event]
 Given an execution $\tau \in \llbracket \M{S}_{\M{P}, \Psi} \rrbracket$ and an event $\eta \in L(\tau)$, $\eta$ is convergent if $\Sigma_{\eta}$ is singleton.
 \end{definition}

\begin{definition}[Strong Eventual Consistency]
A CRDT $(\Sigma, O, \sigma_{\F{init}})$ achieves strong eventual consistency {\sf (SEC)}under a weak consistency specification $\Psi$ if for all well-formed executions $\tau \in \M{WF}(\M{S}_{\M{P}, \Psi})$ and for all events $\eta \in L(\tau)$, $\eta$ is convergent.
\end{definition}

An event is convergent if all valid permutations of visible events
according to the specification $\Psi$ lead to the same state. This
corresponds to the requirement that if two replicas have witnessed the
same set of operations, they must be in the same state. A CRDT
achieves {\sf SEC} if all events in all executions are convergent.


\nocite{BU14,GO16}

\section{Automated Verification}
\label{sec:proofrule}

In order to show that a CRDT achieves {\sf SEC} under a consistency
specification, we need to show that all events in any well-formed execution are convergent, which in turn
requires us to show that any valid permutation of valid subsets of
events in an execution leads to the same state.  This is a hard
problem because we have to reason about executions of unbounded
length, involving unbounded sets of effectors and reconcile the
declarative event-based specifications of weak consistency with states
generated during execution. To make the problem tractable, we use a
two-fold strategy. First, we show that if any pair of effectors
generated during any execution either commute with each other or are
forced to be applied in the same order by the consistency policy, then
the CRDT achieves {\sf SEC}. Second, we develop an inductive proof
rule to show that \emph{all} pairs of effectors generated during any
(potentially unbounded) execution obey the above mentioned property.
To ensure soundness of the proof rule, we place some reasonable
assumptions on the consistency policy that (intuitively) requires
behaviorally equivalent events to be treated the same by the policy,
regardless of context (i.e., the length of the execution history at
the time the event is applied).  We then extract a simple sufficient
condition which we call as \emph{non-interference to commutativity}
that captures the heart of the inductive argument.  Notably, this
condition can be automatically checked for different CRDTs under
different consistency policies using off-the-shelf theorem provers,
thus providing a pathway to performing automated parametrized
verification of CRDTs.

Given a transition $(\Delta, \F{vis}, \F{eo}) \xrightarrow{\eta} C$, we
denote the global effector order in the starting configuration of
$\eta$, i.e. $\F{eo}$ as $\F{eo}_{\eta}$. We first show that a
sufficient condition to prove that a CRDT is convergent is to show that
any two events in its history either commute or are related by the
global effector order.

\begin{lemma}
Given an execution $\tau \in \llbracket \M{S}_{\M{P}, \Psi}
\rrbracket$, and an event $\eta =(\F{id},o,\sigma_s,\Delta_r,\F{eo}_r) \in L(\tau)$ 
, if for all $\eta_1, \eta_2 \in \Delta_r$ such that $\eta_1
\neq \eta_2$, either $\eta_1^e \circ \eta_2^e = \eta_2^e \circ
\eta_1^e$ or $\F{eo}_{\eta}(\eta_1, \eta_2)$ or $\F{eo}_{\eta}(\eta_2,
\eta_1)$, then $\eta$ is convergent \footnote{All proofs can be found in the Appendix A}.
\end{lemma}

We now present a property that consistency policies must obey for our
verification methodology to be soundly applied. First, we define the
notion of behavioral equivalence of events:

\begin{definition}[Behavioral Equivalence]

Two events $\eta_1 = (\F{id}_1, o_1, \sigma_1, \Delta_1, \F{eo}_1)$ and $\eta_2
= (\F{id}_2, o_2, \sigma_2, \Delta_2, \F{eo}_2)$ are behaviorally equivalent if
$\eta_1^{e} = \eta_2^{e}$ and $o_1 = o_2$.
\end{definition}

That is, behaviorally equivalent events produce the same effectors.  We
use the notation $\eta_1 \equiv \eta_2$ to indicate that they are
behaviorally equivalent.

\begin{definition}[Behaviorally Stable Consistency Policy]
A consistency policy $\Psi$ is behaviorally stable if $\forall \Delta, \F{vis}, \F{eo},  \Delta', \F{vis}^{'}, \F{eo}^{'}$, $ \eta_1, \eta_2 \in \Delta$, $\eta_1^{'}, \eta_2^{'} \in \Delta^{'}$ the following holds: 
\begin{equation}
\begin{split}
(\Psi(\Delta, \F{vis}, \F{eo}) \wedge \Psi(\Delta^{'}, \F{vis}^{'}, \F{eo}^{'}) \wedge \eta_1 \equiv \eta_{1}^{'} \wedge \eta_2 \equiv \eta_{2}^{'} \wedge \F{vis}(\eta_1, \eta_2) \Leftrightarrow \F{vis}'(\eta_{1}^{'}, \eta_{2}^{'})) \\
 \Rightarrow  \F{eo}(\eta_1, \eta_2) \Leftrightarrow \F{eo}'(\eta_{1}^{'}, \eta_{2}^{'})
\end{split}
\end{equation}
\end{definition}

Behaviorally stable consistency policies treat behaviorally equivalent
events which have the same visibility relation among them in the same
manner by enforcing the same effector order. All consistency policies
that we discussed in the previous section (representing the most
well-known in the literature) are behaviorally stable:

\begin{lemma}
{\sf EC}, {\sf CC}, {\sf PSI}, {\sf RB} and {\sf SC} are behaviorally stable.
\end{lemma}

{\sf EC} does not enforce any effector ordering and hence is trivially
stable behaviorally. {\sf CC} forces causally dependent events to be
in the same order, and hence behaviorally equivalent events which have
the same visibility order will be forced to be in the same effector
order. {\sf RB} forces events whose operations belong to a specific
subset to be in the same order, but since behaviorally equivalent
events perform the same operation, they would be enforced in the same
effector ordering. Similarly, {\sf PSI} forces events writing to a
common variable to be in the same order, but since behaviorally
equivalent events generate the same effector, they would also write to
the same variables and hence would be forced in the same effector
order. {\sf SC} forces all events to be in the same order which is
equal to the visibility order, and hence is trivially stable
behaviorally. In general, behaviorally stable consistency policies do
not consider the context in which events occur, but instead rely only
on observable behavior of the events to constrain their ordering. A
simple example of a consistency policy which is not behaviorally
stable is a policy which maintains bounded concurrency \cite{KA18} by
limiting the number of concurrent operations across all replicas to a
fixed bound. Such a policy would synchronize two events only if they
occur in a context where keeping them concurrent would violate the
bound, but behaviorally equivalent events in a different context may
not be synchronized.  

%
%
%
%

For executions under a behaviorally stable consistency policy, the
global effector order between events only grows in an execution, so
that if two events $\eta_1$ and $\eta_2$ are in the history of some event
$\eta$ are related by $\F{eo}_{\eta}$, then if they later occur in the
history of any other event, they would be related in the same effector
order. Hence, we can now define a common global effector order for an
execution. Given an execution $\tau \in \llbracket \M{S}_{\M{P},
  \Psi}\rrbracket$, the effector order $\F{eo}_{\tau} \subseteq
L(\tau) \times L(\tau)$ is an anti-symmetric relation defined as
follows:

$$
\F{eo}_{\tau} = \{(\eta_1, \eta_2)\ |\ \exists \eta \in L(\tau).\ (\eta_1, \eta_2) \in \F{eo}_{\eta} \}
$$

Similarly, we also define $\F{vis}_{\tau}$ to be the common visibility relation for an execution $\tau$, which is nothing but the $\F{vis}$ relation in the final configuration of $\tau$.

\begin{definition}[Commutative modulo Consistency Policy]
Given a CRDT $\M{P}$, a behaviorally stable weak consistency specification $\Psi$ and an execution $\tau \in \llbracket \M{S}_{\M{P}, \Psi} \rrbracket$, two events $\eta_1,\eta_2 \in L(\tau)$ such that $\eta_1 \neq \eta_2$ commute modulo the consistency policy $\Psi$ if either $\eta_1^e \circ \eta_2^e = \eta_2^e \circ \eta_1^e$ or $\F{eo}_{\tau}(\eta_1, \eta_2)$ or $\F{eo}_{\tau}(\eta_2, \eta_1)$.
\end{definition}

\noindent The following lemma is a direct consequence of Lemma 1 :

\begin{lemma}
Given a CRDT $\M{P}$ and a behaviorally stable consistency specification $\Psi$, if for all $\tau \in \M{WF}(\M{S}_{\M{P}, \Psi})$, for all $\eta_1, \eta_2 \in L(\tau)$ such that $\eta_1 \neq \eta_2$, $\eta_1$ and $\eta_2$ commute modulo the consistency policy $\Psi$, then $\M{P}$ achieves SEC under $\Psi$.
\end{lemma}

Our goal is to use Lemma 3 to show that all events in any execution
commute modulo the consistency policy. However, executions can be
arbitrarily long and have an unbounded number of events. Hence, for
events occurring in such large executions, we will instead consider
behaviorally equivalent events in a smaller execution and show that
they commute modulo the consistency policy, which by stability of the
consistency policy directly translates to their commutativity in the
larger context.  Recall that the effector generated by an operation
depends on its start state and the sequence of other effectors applied
to that state. To generate behaviorally equivalent events with
arbitrarily long histories in short executions, we summarize these
long histories into the start state of events, and use commutativity
itself as an inductive property of these start states. That is, we ask
if two events with arbitrary start states and empty histories commute
modulo $\Psi$, whether the addition of another event to their
histories would continue to allow them to commute modulo $\Psi$.

\begin{definition}[Non-interference to Commutativity] $(${\sf Non-Interf}$)$
A CRDT $\M{P} = (\Sigma, O, \sigma_{\C{init}})$ satisfies non-interference to commutativity under a consistency policy $\Psi$ if and only if the following conditions hold:
\begin{enumerate}
\item For all executions $C_{\C{init}} \xrightarrow{\eta_1} C_1
  \xrightarrow{\eta_2} C_2$ in $\M{WF}(\M{S}_{\M{P}, \Psi})$,
   $\eta_1$ and $\eta_2$ commute modulo $\Psi$.
\item For all $\sigma_1, \sigma_2, \sigma_3 \in \Sigma$, if for
  execution $\tau \equiv C_{\C{init}} \xrightarrow{\eta_1} C_1
  \xrightarrow{\eta_2} C_2$ in $\llbracket \M{S}_{\M{P}, \Psi} \rrbracket$ where
  $\sigma_s(\eta_1) = \sigma_1$, $\sigma_s(\eta_2) = \sigma_2$,
  $\eta_1$ and $\eta_2$ commute modulo $\Psi$, then for all executions
  $\tau' \equiv C_{\C{init}} \xrightarrow{\eta_3} C_{1}^{'}
  \xrightarrow{\eta_{1}^{'}} C_{2}^{'} \xrightarrow{\eta_{2}^{'}}
  C_{3}^{'}$ such that $\sigma_s(\eta_{1}^{'}) = \sigma_1$, $o(\eta_{1}^{'}) = o(\eta_1)$,
  $\sigma_s(\eta_{2}^{'}) = \sigma_2$, $o(\eta_{2}^{'}) = o(\eta_2)$, $\sigma_s(\eta_3) = \sigma_3$,
  and $\F{vis}_{\tau}(\eta_1, \eta_2) \Leftrightarrow
  \F{vis}_{\tau'}(\eta_{1}^{'}, \eta_{2}^{'})$, $\eta_{1}^{'}$ and
  $\eta_{2}^{'}$ commute modulo $\Psi$.
\end{enumerate}
\end{definition}

Condition (1) corresponds to the base case of our inductive argument
and requires that in well-formed executions with 2 events, both the
events commute modulo $\Psi$. For condition (2), our intention is to
consider two events $\eta_a$ and $\eta_b$ with any arbitrary histories
which can occur in any well-formed execution and, assuming that they
commute modulo $\Psi$, show that even after the addition of another event
to their histories, they continue to commute. We use CRDT states
$\sigma_1,\sigma_2$ to summarize the histories of the two events, and
construct behaviorally equivalent events ($\eta_1 \equiv \eta_a$ and
$\eta_2 \equiv \eta_b$) which would take $\sigma_1,\sigma_2$ as their
start states. That is, if $\eta_a$ produced the effector
$o(\sigma_{\C{init}}, \pi)$\footnote{Note that in a well-formed
  execution, the start state is always $\sigma_{\C{init}}$}, where $o$
is the CRDT operation corresponding to $\eta_a$ and $\pi$ is the
sequence of effectors in its history, we leverage the observation that
$o(\sigma_{\C{init}}, \pi) = o(\pi(\sigma_{\C{init}}), \epsilon)$, and
assuming $\sigma_1 = \pi(\sigma_{\C{init}})$, we obtain the
behaviorally equivalent event $\eta_1$, i.e. $\eta_1^e \equiv
\eta_a^e$. Similar analysis establishes that $\eta_2^e \equiv
\eta_b^e$. However, since we have no way of characterizing states
$\sigma_1$ and $\sigma_2$ which are obtained by applying arbitrary
sequences of effectors, we use commutativity itself as an identifying
characteristic, focusing on only those $\sigma_1$ and $\sigma_2$ for
which the events $\eta_1$ and $\eta_2$ commute modulo $\Psi$.

The interfering event is also summarized by another CRDT state
$\sigma_3$, and we require that after suffering interference from this
new event, the original two events would continue to commute modulo
$\Psi$. This would essentially establish that any two events with any
history would commute modulo $\Psi$ in these small executions, which
by the behavioral stability of $\Psi$ would translate to their
commutativity in any execution.

\begin{theorem}
Given a CRDT $\M{P}$ and a behaviorally stable consistency policy
$\Psi$, if $\M{P}$ satisfies non-interference to commutativity under
$\Psi$, then $\M{P}$ achieves {\sf SEC} under $\Psi$.
\end{theorem}

\textbf{Example:} Let us apply the proposed verification strategy to
the {\sf ORSet} CRDT shown in Fig \ref{fig:ex(b)}. Under {\sf EC}, condition (1)
of {\sf Non-Interf} fails, because in the execution $C_{\C{init}}
\xrightarrow{\eta_1} C_1 \xrightarrow{\eta_2} C_2$ where $o(\eta_1) =
$\C{Add(a,i)} and $o(\eta_2)=$\C{Remove(a)} and $\F{vis}(\eta_1,
\eta_2)$, $\eta_1$ and $\eta_2$ don't commute modulo {\sf EC}, since
\C{(a,i)} would be present in the source replica of
\C{Remove(a)}.  However, $\eta_1$ and $\eta_2$ would commute
modulo {\sf CC}, since they would be related by the effector order. Now,
moving to condition (2) of {\sf Non-interf}, we limit ourselves to
source replica states $\sigma_1$ and $\sigma_2$ where \C{Add(a,i)} and
\C{Remove(a)} do commute modulo {\sf CC}. If $\F{vis}_{\tau}(\eta_1,
\eta_2)$, then after interference, in execution $\tau'$,
$\F{vis}_{\tau'}(\eta_1^{'}, \eta_2^{'})$, in which case $\eta_1^{'}$
and $\eta_2^{'}$ trivially commute modulo {\sf CC} (because they would be
related by the effector order). On the other hand, if $\neg
\F{vis}_{\tau}(\eta_1, \eta_2)$, then for $\eta_1$ and $\eta_2$ to
commute modulo {\sf CC}, we must have that the effectors $\eta_1^e$ and
$\eta_2^e$ themselves commute, which implies that $\C{(a,i)} \notin
\sigma_2$. Now, consider any execution $\tau^{'}$ with an interfering
operation $\eta_3$. If $\eta_3$ is another \C{Add(a,i')} operation,
then $\C{i'}\neq\C{i}$, so that even if it is visible to
$\eta_2^{'}$, $\eta_2^{'e}$ will not remove \C{(a,i)}, so that
$\eta_1^{'}$ and $\eta_2^{'}$ would commute. Similarly, if $\eta_3$ is
another \C{Remove(a)} operation, it can only remove tagged versions of
\C{a} from the source replicas of $\eta_2^{'}$, so that the effector
$\eta_2^{'e}$ would not remove \C{(a,i)}.

\section{Experimental Results}
\label{sec:impl}
  
In this section, we present the results of applying our verification
methodology to a number of CRDTs under different consistency
models.  We collected CRDT implementations
from a number of sources \cite{SH11a,AT16,PR18} and since all of the
existing implementations assume a very weak consistency model
(primarily {\sf CC}), we additionally implemented a few CRDTs on our own
intended to only work under stronger consistency schemes but which are
better in terms of time/space complexity and ease of development.  Our
implementations are not written in any specific language but instead
are specified abstractly akin to the definitions given in
Figs.~\ref{fig:ex(a)} and ~\ref{fig:ex(b)}.  To specify CRDT states
and operations, we fix an abstract language that contains
uninterpreted datatypes (used for specifying elements of sets, lists,
etc.), a set datatype with support for various set operations (add,
delete, union, intersection, projection, lookup), a tuple datatype
(along with operations to create tuples and project components) and a
special uninterpreted datatype equipped with a total order for
identifiers.  Note that the set datatype used in our abstract language
is different from the {\sf Set} CRDT, as it is only intended to
perform set operations locally at a replica. Our data types are
uninterpreted since our verification technique generates verification
conditions in first-order logic (FOL) with universal quantifiers.  All
existing CRDT definitions can be naturally expressed in this
framework and can be found in Appendix B.

Here, we revert back to the op-based specification of CRDTs. For a
given CRDT $\M{P}=(\Sigma,O,\sigma_{\C{init}})$, we convert all its
operations into FOL formulas relating the source, input and output
replica states. That is, for a CRDT operation $o : \Sigma \rightarrow
\Sigma \rightarrow \Sigma$, we create a predicate $\F{o} : \Sigma
\times \Sigma \times \Sigma \rightarrow \mathbb{B}$ such that
$\F{o}(\sigma_s, \sigma_i, \sigma_o)$ is true if and only if
$o(\sigma_s)(\sigma_i) = \sigma_o$. Since CRDT states are typically
expressed as sets, we axiomatize set operations to express their
semantics in FOL.

%

In order to specify a consistency model, we introduce a sort for events
and binary predicates $\F{vis}$ and $\F{eo}$ over this sort. Here, we
can take advantage of the declarative specification of consistency
models and directly encode them in FOL. Given an encoding of CRDT
operations and a consistency model, our verification strategy is to
determine whether the {\sf Non-Interf} property holds. Since both
conditions of this property only involve executions of finite length
(at most 3), we can directly encode them as $\F{UNSAT}$ queries by
asking for executions which break the conditions. For condition (1),
we query for the existence of two events $\eta_1$ and $\eta_2$ along with
$\F{vis}$ and $\F{eo}$ predicates which satisfy the consistency
specification $\Psi$ such that these events are not related by
$\F{eo}$ and their effectors do not commute. For condition (2), we
query for the existence of events $\eta_1, \eta_2, \eta_3$ and their
respective start states $\sigma_1,\sigma_2, \sigma_3$, such that
$\eta_1$ and $\eta_2$ commute modulo $\Psi$ but after interference
from $\eta_3$, they are not related by $\F{eo}$ and do not
commute. Both these queries are encoded in EPR~\cite{PMB10}, a decidable fragment
of FOL, so if the CRDT operations and the consistency policy can also
be encoded in a decidable fragment of FOL (which is the case in all
our experiments), then our verification strategy is also decidable. We
write {\sf Non-Interf-1} and {\sf Non-Interf-2} for the two
conditions of {\sf Non-Interf}.

\begin{figure}
\begin{tabularx}{\textwidth}{ | X | C | C | C | C | C |}
\hline
CRDT & {\sf EC} & {\sf CC} & {\sf PSI+RB} & {\sf PSI} & Verif. Time (s) \\ \hline
\multicolumn{6}{| c |}{\sf Set} \\ \hline
{\sf Simple-Set} & \xmark & \xmark & \cmark & \cmark & 0.23\\ \hline
{\sf ORSet} \cite{SH11a} & \xmark & \cmark & \cmark & \cmark & 0.6\\ \hline
{\sf ORSet} with Tombstones & \cmark & \cmark & \cmark & \cmark & 0.04\\ \hline
{\sf USet}\cite{SH11a} &  \xmark & \xmark & \xmark & \cmark & 0.1\\ \hline
\multicolumn{6}{| c |}{\sf List} \\ \hline
{\sf RGA}\cite{AT16} & \xmark & \cmark & \cmark & \cmark & 5.3\\ \hline
{\sf RGA-No-Tomb} & \xmark & \xmark & \cmark & \cmark & 3 \\ \hline
\multicolumn{6}{| c |}{\sf Graph} \\ \hline
{\sf 2P2P-Graph}\cite{SH11a} & \xmark & \cmark & \cmark & \cmark & 3.5\\ \hline
{\sf Graph-with-ORSet} &  \xmark & \xmark & \cmark & \cmark & 46.3\\ \hline
\end{tabularx}
\caption{Convergence of CRDTs under different consistency policies.}
\label{fig:tab}
\vspace*{-.4in}
\end{figure}

Fig.~\ref{fig:tab} shows the results of applying the proposed
methodology on different CRDTs. We used Z3 to discharge our
satisfiability queries. For every combination of a CRDT and a
consistency policy, we write \xmark\ to indicate that verification of
{\sf Non-Interf} failed, while \cmark\ indicates that it was
satisfied. We also report the verification time taken by Z3 for every
CRDT across all consistency policies executing on a standard desktop
machine.  We have picked the three collection datatypes for which
CRDTs have been proposed i.e. {\sf Set}, {\sf List} and {\sf Graph},
and for each such datatype, we consider multiple variants that provide
a tradeoff between consistency requirements and implementation
complexity. Apart from {\sf EC}, {\sf CC} and {\sf PSI}, we also use a
combination of {\sf PSI} and {\sf RB}, which only enforce {\sf PSI}
between selected pairs of operations (in contrast to simple {\sf RB}
which would enforce {\sf SC} between all selected pairs). Note that
when verifying a CRDT under {\sf PSI}, we assume that the set
operations are implemented as Boolean assignments, and the write set
$\F{Wr}$ consists of elements added/removed.  We are unaware of any
prior effort that has been successful in automatically verifying
\emph{any} CRDT, let alone those that exhibit the complexity of the
ones considered here.

\textbf{Set}: The {\sf Simple-Set} CRDT in Fig \ref{fig:ex(a)} does
not converge under {\sf EC} or {\sf CC}, but achieves
  convergence under {\sf PSI+RB} which only synchronizes \C{Add} and \C{Remove} operations to the same
  elements, while all other operations continue to run under {\sf EC}, since they do commute
  with each other.
   As explained earlier, {\sf ORSet} does not
  converge under {\sf EC} and violates {\sf Non-Interf-1}. {\sf ORSet with
  tombstones} converges under {\sf EC} as well since it uses a different set
  (called a tombstone) to keep track of removed elements. {\sf USet} is
  another implementation of the Set CRDT which converges under the
  assumptions that an element is only added once, and removes only
  work if the element is already present in the source replica. {\sf USet}
  converges only under {\sf PSI}, because under any weaker consistency
  model, {\sc non-interf-2} breaks, since \C{Add(a)} interferes and
  breaks the commutativity of \C{Add(a)} and \C{Remove(a)}. Notice
  that as the consistency level weakens, implementations need to keep
  more and more information to maintain convergence--compute unique
  ids, tag elements with them or keep track of deleted elements. If
  the underlying replicated store supports stronger consistency levels
  such as {\sf PSI}, simpler definitions are sufficient.

\textbf{List}: The List CRDT maintains a total ordering between its
elements. It supports two operations : \C{AddRight(e,a)} adds new
element \C{a} to the right of existing element \C{e}, while
\C{Remove(e)} removes \C{e} from the list. We use the implementation
in \cite{AT16} (called RGA) which uses time-stamped insertion
trees. To maintain integrity of the tree structure, the immediate
predecessor of every list element must be present in the list, due to
which operations \C{AddRight(a,b)} and \C{AddRight(b,c)} do not
commute. Hence {\sf RGA} does not converge under {\sf EC} because {\sf Non-Interf-1}
is violated , but converges under {\sf CC}. 

To make adds and removes involving the same list element commute, {\sf
  RGA} maintains a tombstone set for all deleted list elements. This
can be expensive as deleted elements may potentially need to be
tracked forever, even with garbage collection. We consider a slight
modification of {\sf RGA} called {\sf RGA-No-Tomb} which does not keep
track of deleted elements. This CRDT now has a convergence violation
under {\sf CC} (because of {\sf Non-Interf-1}), but achieves
convergence under {\sf PSI+RB} where we enforce {\sf PSI} only for
pairs of \C{AddRight} and \C{Remove} operations.

\textbf{Graph}: The {\sf Graph} CRDT maintains sets of vertices and
edges and supports operations to add and remove vertices and
edges. The {\sf 2P2P-Graph} specification uses separate {\sf 2P-Sets}
for both vertices and edges, where a {\sf 2P-Set} itself maintains two
sets for addition and removal of elements. While {\sf 2P} sets
themselves converge under {\sf EC}, the {\sf 2P2P-Graph} has
convergence violations (to {\sf Non-Interf-1}) involving
\C{AddVertex(v)} and \C{RemoveVertex(v)} (similarly for edges) since
it removes a vertex from a replica only if it is already present.  We
verify that it converges under {\sf CC}. Graphs require an integrity
constraint that edges in the edge-set must always be incident on
vertices in the vertex-set. Since concurrent \C{RemoveVertex(v)} and
\C{AddEdge(v,v')} can violate this constraint, the {\sf 2P2P-Graph}
uses the internal structure of the {\sf 2P-Set} which keeps track of
deleted elements and considers an edge to be in the edge set only if
its vertices are not in the vertex tombstone set (leading to a remove-wins
strategy).

Building a graph CRDT can be viewed as an exercise in composing CRDTs
by using two {\sf ORSet} CRDTs, keeping the internal implementation of
the {\sf ORSet} opaque, using only its interface. The {\sf
  Graph-with-ORSet} implementation uses separate {\sf ORSet}s for
vertices and edges and explicitly maintains the graph integrity
constraint. We find convergence violations (to {\sf Non-Interf-1})
between \C{RemoveVertex(v)} and \C{AddEdge(v,v')}, and
\C{RemoveVertex(v)} and \C{RemoveEdge(v,v')} under both {\sf EC} and
  {\sf CC}. Under {\sf PSI+RB} (enforcing {\sf RB} on the above two
  pairs of operations), we were able to show convergence.

When a CRDT passes {\sf Non-Interf} under a consistency policy, we can
guarantee that it achieves {\sf SEC} under that policy. However, if it
fails {\sf Non-Interf}, it may or may not converge. In particular, if
it fails {\sf Non-Interf-1} it will definitely not converge (because
{\sf Non-Interf-1} constructs a well-formed execution), but if it
passes {\sf Non-Interf-1} and fails {\sf Non-Interf-2}, it may still
converge because of the imprecision of {\sf Non-Interf-2}. There are
two sources of imprecision, both concerning the start states of the
events picked in the condition: (1) we only use commutativity as a
distinguishing property of the start states, but this may not be a
sufficiently strong inductive invariant, (2) we place no constraints
on the start state of the interfering operation.  In practice, we have
found that for all cases except {\sf U-Set}, convergence violations
manifest via failure of {\sf Non-Interf-1}.  If {\sf Non-Interf-2}
breaks, we can search for well-formed executions of higher length
upto a bound. For {\sf U-Set}, we were successful in adopting this
approach, and were able to find a non-convergent well-formed
execution of length 3.

\section{Related Work and Conclusions}
\label{sec:related}

Reconciling concurrent updates in a replicated system is a important
well-studied problem in distributed applications, having been first
studied in the context of collaborative editing systems
\cite{NC95}. Incorrect implementation of replicated sets in Amazon's
Dynamo system \cite{DH07} motivated the design of CRDTs as a
principled approach to implementing replicated data types.  Devising
correct implementations has proven to be challenging, however, as
evidenced by the myriad pre-conditions specified in the various CRDT
implementations \cite{SH11a}.

Burckhardt \emph{et al.}~\cite{BU14} present an abstract event-based
framework to describe executions of CRDTs under different network
conditions; they also propose a rigorous correctness criterion in the
form of abstract specifications. Their proof strategy, which is
neither automated nor parametric on consistency policies, verifies
CRDT implementations against these specifications by providing a
simulation invariant between CRDT states and event structures.  Zeller
\emph{et al.}~\cite{ZE14} also require simulation invariants to verify
convergence, although they only target state-based CRDTs.  Gomes
\emph{et al.}~\cite{GO17} provide mechanized proofs of convergence for
{\sf ORSet} and {\sf RGA} CRDTs under causal consistency, but
their approach is neither automated nor parametric.

A number of earlier efforts~\cite{BA14,SI15,GO16,KA18,HO19} have
looked at the problem of verifying state-based invariants in
distributed applications.  These techniques typically target
applications built using CRDTs, and assume their underlying
correctness. Because they target correctness specifications in the
form of state-based invariants, it is unclear if their approaches can
be applied directly to the convergence problem we consider here.
Other approaches \cite{BG16,BR18,NJ18} have also looked at the
verification problem of transactional programs running on replicated
systems under weak consistency, but these proposals typically use
serializability as the correctness criterion, adopting a ``last-writer
wins'' semantics, rather than convergence, to deal with concurrent
updates.

This paper demonstrates the automated verification of CRDTs under
different weak consistency policies.  We rigorously define the
relationship between commutativity and convergence, formulating the
notion of commutativity modulo consistency policy as a sufficient
condition for convergence.  While we require a non-trivial inductive
argument to show that non-interference to commutativity is sufficient
for convergence, the condition itself is designed to be simple and
amenable to automated verification using off-the-shelf
theorem-provers. We have successfully applied the proposed
verification strategy for all major CRDTs, additionally motivating the
need for parametrization in consistency policies by showing variants
of existing CRDTs which are simpler in terms of implementation
complexity but converge under different weak consistency models.

\paragraph{Acknowledgments.}  We thank the anonymous reviewers for their
  insightful comments.  This material is based upon work supported by
  the National Science Foundation under under Grant No. CCF-SHF
  1717741 and the Air Force Research Lab under Grant No.
  FA8750-17-1-0006.

\bibliographystyle{splncs04}
\bibliography{db}

\begin{thebibliography}{10}
\providecommand{\url}[1]{\texttt{#1}}
\providecommand{\urlprefix}{URL }
\providecommand{\doi}[1]{https://doi.org/#1}

\bibitem{AT16}
Attiya, H., Burckhardt, S., Gotsman, A., Morrison, A., Yang, H., Zawirski, M.:
  Specification and complexity of collaborative text editing. In: Proceedings
  of the 2016 {ACM} Symposium on Principles of Distributed Computing, {PODC}
  2016, Chicago, IL, USA, July 25-28, 2016. pp. 259--268 (2016).
  \doi{10.1145/2933057.2933090}, \url{https://doi.org/10.1145/2933057.2933090}

\bibitem{BA14}
Bailis, P., Fekete, A., Franklin, M.J., Ghodsi, A., Hellerstein, J.M., Stoica,
  I.: Coordination avoidance in database systems. {PVLDB}  \textbf{8}(3),
  185--196 (2014). \doi{10.14778/2735508.2735509},
  \url{http://www.vldb.org/pvldb/vol8/p185-bailis.pdf}

\bibitem{BA13}
Bailis, P., Ghodsi, A.: Eventual consistency today: limitations, extensions,
  and beyond. Commun. {ACM}  \textbf{56}(5),  55--63 (2013).
  \doi{10.1145/2447976.2447992}, \url{https://doi.org/10.1145/2447976.2447992}

\bibitem{BG16}
Bernardi, G., Gotsman, A.: Robustness against consistency models with atomic
  visibility. In: 27th International Conference on Concurrency Theory, {CONCUR}
  2016, August 23-26, 2016, Qu{\'{e}}bec City, Canada. pp. 7:1--7:15 (2016).
  \doi{10.4230/LIPIcs.CONCUR.2016.7},
  \url{https://doi.org/10.4230/LIPIcs.CONCUR.2016.7}

\bibitem{BR18}
Brutschy, L., Dimitrov, D., M{\"{u}}ller, P., Vechev, M.T.: Static
  serializability analysis for causal consistency. In: Proceedings of the 39th
  {ACM} {SIGPLAN} Conference on Programming Language Design and Implementation,
  {PLDI} 2018, Philadelphia, PA, USA, June 18-22, 2018. pp. 90--104 (2018).
  \doi{10.1145/3192366.3192415}, \url{https://doi.org/10.1145/3192366.3192415}

\bibitem{BU14}
Burckhardt, S., Gotsman, A., Yang, H., Zawirski, M.: Replicated data types:
  specification, verification, optimality. In: The 41st Annual {ACM}
  {SIGPLAN-SIGACT} Symposium on Principles of Programming Languages, {POPL}
  '14, San Diego, CA, USA, January 20-21, 2014. pp. 271--284 (2014).
  \doi{10.1145/2535838.2535848}, \url{https://doi.org/10.1145/2535838.2535848}

\bibitem{DH07}
DeCandia, G., Hastorun, D., Jampani, M., Kakulapati, G., Lakshman, A., Pilchin,
  A., Sivasubramanian, S., Vosshall, P., Vogels, W.: Dynamo: amazon's highly
  available key-value store. In: Proceedings of the 21st {ACM} Symposium on
  Operating Systems Principles 2007, {SOSP} 2007, Stevenson, Washington, USA,
  October 14-17, 2007. pp. 205--220 (2007). \doi{10.1145/1294261.1294281},
  \url{https://doi.org/10.1145/1294261.1294281}

\bibitem{GI02}
Gilbert, S., Lynch, N.A.: Brewer's conjecture and the feasibility of
  consistent, available, partition-tolerant web services. {SIGACT} News
  \textbf{33}(2),  51--59 (2002). \doi{10.1145/564585.564601},
  \url{http://doi.acm.org/10.1145/564585.564601}

\bibitem{GO17}
Gomes, V.B.F., Kleppmann, M., Mulligan, D.P., Beresford, A.R.: Verifying strong
  eventual consistency in distributed systems. {PACMPL}  \textbf{1}({OOPSLA}),
  109:1--109:28 (2017). \doi{10.1145/3133933},
  \url{https://doi.org/10.1145/3133933}

\bibitem{GO16}
Gotsman, A., Yang, H., Ferreira, C., Najafzadeh, M., Shapiro, M.: 'cause i'm
  strong enough: reasoning about consistency choices in distributed systems.
  In: Proceedings of the 43rd Annual {ACM} {SIGPLAN-SIGACT} Symposium on
  Principles of Programming Languages, {POPL} 2016, St. Petersburg, FL, USA,
  January 20 - 22, 2016. pp. 371--384 (2016). \doi{10.1145/2837614.2837625},
  \url{http://doi.acm.org/10.1145/2837614.2837625}

\bibitem{HO19}
Houshmand, F., Lesani, M.: Hamsaz: replication coordination analysis and
  synthesis. {PACMPL}  \textbf{3}({POPL}),  74:1--74:32 (2019),
  \url{https://dl.acm.org/citation.cfm?id=3290387}

\bibitem{KA18}
Kaki, G., Earanky, K., Sivaramakrishnan, K.C., Jagannathan, S.: Safe
  replication through bounded concurrency verification. {PACMPL}
  \textbf{2}({OOPSLA}),  164:1--164:27 (2018). \doi{10.1145/3276534},
  \url{https://doi.org/10.1145/3276534}

\bibitem{LI12}
Li, C., Porto, D., Clement, A., Gehrke, J., Pregui{\c{c}}a, N.M., Rodrigues,
  R.: Making geo-replicated systems fast as possible, consistent when
  necessary. In: 10th {USENIX} Symposium on Operating Systems Design and
  Implementation, {OSDI} 2012, Hollywood, CA, USA, October 8-10, 2012. pp.
  265--278 (2012),
  \url{https://www.usenix.org/conference/osdi12/technical-sessions/presentation/li}

\bibitem{LL11}
Lloyd, W., Freedman, M.J., Kaminsky, M., Andersen, D.G.: Don't settle for
  eventual: scalable causal consistency for wide-area storage with {COPS}. In:
  Proceedings of the 23rd {ACM} Symposium on Operating Systems Principles 2011,
  {SOSP} 2011, Cascais, Portugal, October 23-26, 2011. pp. 401--416 (2011).
  \doi{10.1145/2043556.2043593},
  \url{http://doi.acm.org/10.1145/2043556.2043593}

\bibitem{NJ18}
Nagar, K., Jagannathan, S.: Automated detection of serializability violations
  under weak consistency. In: 29th International Conference on Concurrency
  Theory, {CONCUR} 2018, September 4-7, 2018, Beijing, China. pp. 41:1--41:18
  (2018). \doi{10.4230/LIPIcs.CONCUR.2018.41},
  \url{https://doi.org/10.4230/LIPIcs.CONCUR.2018.41}

\bibitem{NC95}
Nichols, D.A., Curtis, P., Dixon, M., Lamping, J.: High-latency, low-bandwidth
  windowing in the jupiter collaboration system. In: Proceedings of the 8th
  Annual {ACM} Symposium on User Interface Software and Technology, {UIST}
  1995, Pittsburgh, PA, USA, November 14-17, 1995. pp. 111--120 (1995).
  \doi{10.1145/215585.215706}, \url{https://doi.org/10.1145/215585.215706}

\bibitem{PMB10}
Piskac, R., de~Moura, L.M., Bj{\o}rner, N.: Deciding effectively propositional
  logic using {DPLL} and substitution sets. J. Autom. Reasoning
  \textbf{44}(4),  401--424 (2010). \doi{10.1007/s10817-009-9161-6},
  \url{https://doi.org/10.1007/s10817-009-9161-6}

\bibitem{PR18}
Pregui{\c{c}}a, N.M., Baquero, C., Shapiro, M.: Conflict-free replicated data
  types (crdts). CoRR  \textbf{abs/1805.06358} (2018),
  \url{http://arxiv.org/abs/1805.06358}

\bibitem{SH11a}
Shapiro, M., Pregui{\c{c}}a, N., Baquero, C., Zawirski, M.: {A comprehensive
  study of Convergent and Commutative Replicated Data Types}. Tech. Rep.
  RR-7506, INRIA, Inria – Centre Paris-Rocquencourt (2011)

\bibitem{SH11b}
Shapiro, M., Pregui{\c{c}}a, N.M., Baquero, C., Zawirski, M.: Conflict-free
  replicated data types. In: Stabilization, Safety, and Security of Distributed
  Systems - 13th International Symposium, {SSS} 2011, Grenoble, France, October
  10-12, 2011. Proceedings. pp. 386--400 (2011).
  \doi{10.1007/978-3-642-24550-3\_29},
  \url{https://doi.org/10.1007/978-3-642-24550-3\_29}

\bibitem{SI15}
Sivaramakrishnan, K.C., Kaki, G., Jagannathan, S.: Declarative programming over
  eventually consistent data stores. In: Proceedings of the 36th {ACM}
  {SIGPLAN} Conference on Programming Language Design and Implementation,
  Portland, OR, USA, June 15-17, 2015. pp. 413--424 (2015).
  \doi{10.1145/2737924.2737981}, \url{https://doi.org/10.1145/2737924.2737981}

\bibitem{SO11}
Sovran, Y., Power, R., Aguilera, M.K., Li, J.: Transactional storage for
  geo-replicated systems. In: Proceedings of the 23rd {ACM} Symposium on
  Operating Systems Principles 2011, {SOSP} 2011, Cascais, Portugal, October
  23-26, 2011. pp. 385--400 (2011). \doi{10.1145/2043556.2043592},
  \url{http://doi.acm.org/10.1145/2043556.2043592}

\bibitem{ZE14}
Zeller, P., Bieniusa, A., Poetzsch{-}Heffter, A.: Formal specification and
  verification of crdts. In: Formal Techniques for Distributed Objects,
  Components, and Systems - 34th {IFIP} {WG} 6.1 International Conference,
  {FORTE} 2014, Held as Part of the 9th International Federated Conference on
  Distributed Computing Techniques, DisCoTec 2014, Berlin, Germany, June 3-5,
  2014. Proceedings. pp. 33--48 (2014). \doi{10.1007/978-3-662-43613-4\_3},
  \url{https://doi.org/10.1007/978-3-662-43613-4\_3}

\end{thebibliography}
\appendix
\section{Proofs}

\noindent \textbf{Lemma 1.} Given an execution $\tau \in \llbracket \M{S}_{\M{P}, \Psi}
\rrbracket$, and an event $\eta =(\F{id},o,\sigma_s, \Delta_r,\F{eo}_r) \in L(\tau)$, 
if for all $\eta_1, \eta_2 \in \Delta_r$ such that $\eta_1
\neq \eta_2$, either $\eta_1^e \circ \eta_2^e = \eta_2^e \circ
\eta_1^e$ or $\F{eo}_{\eta}(\eta_1, \eta_2)$ or $\F{eo}_{\eta}(\eta_2,
\eta_1)$, then $\eta$ is convergent.

\begin{proof}
Let $\Delta_r = \{\eta_1, \ldots, \eta_n\}$. To show that $\eta$ is convergent, we need to show that any two permutations $P_1$ and $P_2$ of effectors in $\Delta_r$ obeying the global effector order $\F{eo}_{\eta}$ lead to the same state when applied on the initial state $\sigma_{\C{init}}$. We will show this using induction on $n$ (i.e. size of $\Delta_r$). If $n=1$, then there is only one effector and hence only one permutation. Assume that the result holds for $k-1$. Now, let $n=k$ and consider two permutations $P_1$ and $P_2$. We need to show that 

$$
\eta_{P_1(1)}^{e} \circ \ldots \eta_{P_1(k)}^{e} (\sigma_{\C{init}})=\eta_{P_2(1)}^{e} \circ \ldots \eta_{P_2(k)}^{e} (\sigma_{\C{init}})
$$

If $P_1(1) = P_2(1)$, then the result follows because by the inductive hypothesis, which states that if the number of events is $k-1$ then all permutations lead to the same state, and hence 

$$
\eta_{P_1(2)}^{e} \circ \ldots \eta_{P_1(k)}^{e} (\sigma_{\C{init}})=\eta_{P_2(2)}^{e} \circ \ldots \eta_{P_2(k)}^{e} (\sigma_{\C{init}})
$$

If $P_1(1) \neq P_2(1)$, let $j$ be such that $P_1(1) = P_2(j)$. Now, for all $1 \leq k < j$, neither $\F{eo}_{\eta}(\eta_{P_1(1)}, \eta_{P_2(k)})$ nor $\F{eo}_{\eta}(\eta_{P_2(k)},\eta_{P_1(1)})$ because $\eta_{P_2(k)}$ appears before $\eta_{P_1(1)}$ in $P_2$ and $\eta_{P_1(1)}$ appears before $\eta_{P_2(k)}$ in $P_1$. Hence, we must have 

$$
\eta_{P_1(1)}^{e} \circ \eta_{P_2(k)}^{e} = \eta_{P_2(k)}^{e} \circ \eta_{P_1(1)}^{e}
$$

Hence, we now interchange $\eta_{P_2(j)}^{e}$ and $\eta_{P_2(k)}^{e}$ without changing the overall function obtained from the composition, bringing $\eta_{P_2(j)}^{e}$ at the front of $P_2$. Now, the result directly follows as above from the inductive hypothesis. \qed
\end{proof}

\noindent \textbf{Lemma 2.} {\sf EC}, {\sf CC}, {\sf PSI}, {\sf RB} and {\sf SC} are behaviorally stable.
\begin{proof}
Directly follows from the definition of behavioral stability and definition of the consistency policies.\qed
\end{proof}

\noindent \textbf{Lemma 3.} Given a CRDT $\M{P}$ and a behaviorally stable consistency specification $\Psi$, if for all $\tau \in \M{WF}(\M{S}_{\M{P}, \Psi})$, for all $\eta_1, \eta_2 \in L(\tau)$ such that $\eta_1 \neq \eta_2$, $\eta_1$ and $\eta_2$ commute modulo the consistency policy $\Psi$, then $\M{P}$ achieves SEC under $\Psi$.

\begin{proof}
Given an execution $\tau \in \M{WF}(\M{S}_{\M{P}, \Psi})$, and an event $\eta \in L(\tau)$, let $\eta = (\F{id}, o, \sigma_{\C{init}}, \Delta_r, \F{eo}_r)$. Consider $\eta_1, \eta_2 \in \Delta_r$. Since $\eta_1, \eta_2 \in L(\tau)$, they commute modulo the consistency policy. By behavioral stability of $\Psi$, we have 

$$
\F{eo}_{\tau}(\eta_1, \eta_2) \Rightarrow \F{eo}_{\eta}(\eta_1, \eta_2)
$$ 

This implies that either $\eta_1^{e} \circ \eta_2^{e}$ or $\F{eo}_{\eta}(\eta_1, \eta_2)$ or $\F{eo}_{\eta}(\eta_2, \eta_1)$. By Lemma 1, we have the fact the $\eta$ is convergent. Hence, by definition of {\sc SEC}, the result directly follows. \qed
\end{proof}

\noindent \textbf{Theorem 1.} Given a CRDT $\M{P}$ and a behaviorally stable consistency policy
$\Psi$, if $\M{P}$ satisfies non-interference to commutativity under
$\Psi$, then $\M{P}$ achieves {\sf SEC} under $\Psi$.

\begin{proof}
To prove this, we will show that if $\M{P}$ satisfies non-interference to commutativity under $\Psi$, then any two events in any well-formed execution commute modulo $\Psi$, which by Lemma 3 shows that $\M{P}$ achieves SEC under $\Psi$. We use induction on the length of execution (note that length of an execution is the number of events it contains).

\noindent \textbf{Base Case:} For execution of length 1, it is trivially true since there is only one event. 

\noindent \textbf{Inductive Case:} Suppose the statement holds for all executions of length less than or equal to $k$. Consider an execution $\tau$ of length $k+1$. 
$$
\tau = C_{\C{init}} \rightarrow \ldots \rightarrow C'' \xrightarrow{\eta'} C' \xrightarrow{\eta} C 
$$
In the above execution, $\eta$ is the $(k+1)$th event, and $\eta^{'}$ is its predecessor event. By the inductive hypothesis, all events in $\tau$ before $\eta$ commute with each other modulo $\Psi$, and hence we only need to show that $\eta$ commutes modulo $\Psi$ with all other events in $\tau$. Let $\eta = (\F{id}, o, \sigma_{\C{init}}, \Delta_r, \F{eo})$. There are two cases based on whether $\eta'$ is present in $\Delta_r$:

\textbf{Case-1 :} $\eta' \notin \Delta_r$. In this case, we can interchange the order of the two events, since the event $\eta$ can also occur from configuration $C''$. By doing so, we get an execution of length $k$ containing $\eta$ and by the inductive hypothesis, $\eta$ would commute with all other events in this execution modulo $\Psi$. Hence, we only need to show now that $\eta'$ commutes with $\eta$ modulo $\Psi$. 

Let $\eta^{'} = (\F{id}^{'}, o^{'},  \sigma_{\C{init}}, \Delta_r^{'}, \F{eo}^{'})$. If $\Delta_r \cup \Delta_{r}^{'}$ is empty, then condition (1) in the {\sc non-interf} directly applies, since both these events can directly occur from the initial empty configuration $C_{\C{init}}$. Hence, $\eta$ and $\eta^{'}$ would commute modulo $\Psi$.

Now suppose that $\Delta_r \cup \Delta_{r}^{'}$ contains at least one event. Let $\eta^{''}$ be the latest event in the execution $\tau$ occurring in $\Delta_r \cup \Delta_{r}^{'}$. We can construct an execution $\tau^{'}$ which does not contain $\eta^{''}$ which would end in events $\gamma'$ and $\gamma$ which are analogous to $\eta'$ and $\eta$ except that they don't contain $\eta^{''}$ in their history. Note that every other event which contains $\eta^{''}$ in its history and other events which contain these removed events in their history would also have to be removed to get $\tau'$. However, since $\eta^{''}$ is the latest event in the histories of $\eta$ and $\eta'$, every other event in their history would still be present in $\tau^{'}$. Hence, $\gamma = (\F{id}, o, \sigma_{\C{init}},\Delta_r \setminus \{\eta^{''}\}, \F{eo}_r)$ and $\gamma^{'} =(\F{id}^{'}, o^{'}, \sigma_{\C{init}}, \Delta_r^{'} \setminus \{\eta^{''}\}, \F{eo}^{'}_r)$, where $\F{eo}_r$ and $\F{eo}^{'}_{r}$ are same as the original local effector orders $\F{eo}$ and $\F{eo}^{'}$ but applied on the new histories.

Since the length of $\tau^{'}$ is less than or equal to $k$ (because at least one event is guaranteed to be removed), by the inductive hypothesis $\gamma^{'}$ and $\gamma$ commute modulo $\Psi$. Now, we will instantiate condition (2) of {\sc non-interf} with $\eta_1$ being analogous to $\gamma^{'}$ and $\eta_2$ analogous to $\gamma$. 

Specifically, let $\sigma_1 = \pi'(\sigma_{\C{init}})$ where $\pi'$ is the sequence of effectors in $\Delta_r^{'} \setminus \{\eta^{''}\}$ obeying the total order $\F{eo}^{'}_{r}$. Let $\eta_1 = (\F{id}^{'}, o_1, \sigma_1, \{\}, \{\})$. Then,

\begin{mathpar}
\begin{array}{lcl}
\gamma^{'e} & = & o'(\sigma_{\C{init}}, \pi')\\
& = & o'(\pi'(\sigma_{\C{init}}), \epsilon)\\
& = & o'(\sigma_1, \epsilon)\\
& = & \eta_{1}^{e}
\end{array}
\end{mathpar}

In the above derivation, we first use the property that for any CRDT operation $o$, state $\sigma$ and function $f$, $o(\sigma, f) = o(f(\sigma), \epsilon)$.

Similarly, let $\sigma_2 = \pi(\sigma_{\C{init}})$ where $\pi$ is the sequence of effectors in $\Delta_r \setminus \{\eta^{''}\}$ obeying the total order $\F{eo}_r$. Let $\eta_2 = (\F{id}, o, \sigma_2, \{\}, \{\})$. Similar to above, we can derive that $\gamma^e = \eta_2^e$. Since both $\gamma^{'}$ and $\eta_1$, $\gamma$ and $\eta_2$ have the same effectors and the same visibility relation, by behavioral stability of $\Psi$, in the execution $C_{\C{init}} \xrightarrow{\eta_1} C_1 \xrightarrow{\eta_2} C_2$, $\eta_1$ and $\eta_2$ commute modulo $\Psi$. 

The interfering operation $\eta_3$ is constructed to be analogous to $\eta^{''}$. That is, $\eta_{3}^{e} = \eta^{''e}$. Then, by {\sc non-interf}, for all executions $C_{\C{init}} \xrightarrow{\eta_3} C_{1}^{'} \xrightarrow{\eta_{1}^{'}} C_{2}^{'} \xrightarrow{\eta_{2}^{'}} C_{3}^{'}$ such that $\sigma_s(\eta_{1}^{'}) = \sigma_1$, $\sigma_s(\eta_{2}^{'}) = \sigma_2$ and $\F{vis}_{\tau}(\eta_1, \eta_2) \Leftrightarrow \F{vis}_{\tau'}{\eta_{1}^{'}, \eta_{2}^{'}}$, $\eta_{1}^{'}$ and $\eta_{2}^{'}$ commute modulo $\Psi$. We pick one such execution depending on whether $\eta^{''}$ was visible to $\eta$ and/or $\eta^{'}$.

Let us consider the case where $\eta^{''}$ was visible to $\eta^{'}$, so that $\eta^{''} \in \Delta_{r}^{'}$. Then, $\eta_{1}^{'} = (\F{id}^{'}, o, \sigma_1, \{\eta_3\},\{\})$. We will now show that $\eta_{1}^{'e} = \eta^{'e}$:

\begin{mathpar}
\begin{array}{lcl}
\eta_{1}^{'e} & = & o'(\sigma_1, \eta_3^e)\\
& = & o'(\sigma_1,  \eta^{''e})\\
& = & o'(\pi'(\sigma_{\C{init}},  \eta^{''e}))\\
& = & o'(\sigma_{\C{init}}, \pi' \circ  \eta^{''e})\\
& = & \eta^{'e}
\end{array}
\end{mathpar} 

Note that for the last step, we use the fact that $\eta^{''}$ commutes with all events in $\Delta_{r}^{'}$ modulo $\Psi$ by the inductive hypothesis. Since $\eta^{''}$ is the latest occurring event in $\tau$, it cannot be related by $\F{eo}^{'}$ before any other event in $\Delta_{r}^{'}$. Hence, $\eta^{''}$ is either the last event in the total order $\F{eo}^{'}$ (in which case the above derivation directly follows) or is not the last event, but commutes with every other event after it in $\pi'$, due to which we get the same result.

Similarly, depending on whether $\eta^{''} \in \Delta_r$, we choose an appropriate $\eta_{2}^{'}$ such that $\eta_{2}^{'e} = \eta^{e}$. Now, since $\eta \equiv \eta_{2}^{'}$ and $\eta' \equiv \eta_{1}^{'}$ and $\eta_{1}^{'}$ and $\eta_{2}^{'}$ commute modulo $\Psi$, by behavioral stability of $\Psi$, $\eta$ and $\eta'$ also commute modulo $\Psi$. 

\textbf{Case-2} : $\eta' \in \Delta_r$, let us first
 consider any other event $\eta^{''}$ in $\tau$. Here, we repeat the above procedure to prove the commutativity of the two events $\eta$ and $\eta^{''}$ by considering $\eta^{'} $ as the interfering event.
 
We construct an execution $\tau'$ which is the same as $\tau$ except that it does not contain $\eta'$. This execution will end in the event $\gamma$ which is analogous to $\eta$ except that it does not contain $\eta'$ in its history. Hence, $\gamma=(\F{id},o,\sigma_{\C{init}}, \Delta_r \setminus \{\eta'\}, \F{eo}_r)$, where $\F{eo}_r$ is $\F{eo}$ restricted to $\Delta_r \setminus \{\eta'\}$. The event $\eta^{''}$ will also be present in $\tau'$.  Since the length of $\tau'$ is $k$, by the inductive hypothesis, $\gamma$ and $\eta^{''}$ commute modulo $\Psi$. Now, we will instantiate {\sf NON-INTERF-2} with $\eta_1 \equiv \eta^{''}$ and $\eta_2 \equiv \gamma$.

Specifically, suppose that $\eta^{''} = (\_,o^{''}, \sigma_{\C{init}}, \Delta^{''}, \F{eo}^{''})$. Then \linebreak
 $\eta_1 = (\_, o^{''}, \sigma_1, \{\}, \{\})$, where $\sigma_1 = \pi''(\sigma_{\C{init}})$, where $\pi''$ is the sequence of effectors of $\Delta^{''}$ in the order $\F{eo}^{''}$. Similarly, $\eta_2 = (\_, o, \sigma_2, \{\}, \{\})$, where $\sigma_2 = \pi(\sigma_{\C{init}})$ where $\pi$ is the sequence of effectors of $\Delta_r \setminus \{\eta^{'}\}$ in the order $\F{eo}_r$. Note that here we assume that $\eta^{''} \notin \Delta_r$. The other case where $\eta^{''} \in \Delta_r$ can be similarly handled by placing $\eta_1$ in the history of $\eta_2$. Now, by the behavioral stability of $\Psi$, in the execution, $C_{\C{init}} \xrightarrow{\eta_1} C_1 \xrightarrow{\eta_2} C_2$, $\eta_1$ and $\eta_2$ commute modulo $\Psi$. 

We let $\eta_3 \equiv \eta^{'}$ be the interfering event. Then, in the execution $C_{\C{init}} \xrightarrow{\eta_3} D_{1} \xrightarrow{\eta_1} D_2 \xrightarrow{\eta^{'}_{2}} D_3$ where $\eta^{'}_2 = (\_,o,\sigma_2,\{\eta_3\}, \{\})$, by {\sf NON-INTERF-2}, $\eta_1$ and $\eta_{2}^{'}$ commute modulo $\Psi$. However, $\eta_{2}^{'} \equiv \eta$ and hence by the behavioral stability of $\Psi$, $\eta^{''}$ and $\eta$ commute modulo $\Psi$.

Finally, let us now prove the commutativity of $\eta$ and $\eta^{'}$, where we will again the repeat the above argument by picking the latest event in the histories of $\eta$ and $\eta'$ as the interfering event. 

Let $\eta^{'} = (\_, o^{'},  \sigma_{\C{init}}, \Delta_r^{'}, \F{eo}^{'})$. First, if $\Delta_r \cup \Delta_r^{'} = \{\eta'\}$, then we can directly apply {\sf NON-INTERF-1} since both these events can directly occur from the initial empty configuration, and hence they would commute modulo $\Psi$.

Otherwise, let $\eta^{''}$ be the latest event in $(\Delta_r \cup \Delta_r^{'}) \setminus \{\eta'\}$. We construct an execution $\tau'$ which is the same as $\tau$ except it does not contain $\eta^{''}$ and ends in events $\gamma'$ and $\gamma$ analogous to $\eta'$ and $\eta$ except that they don't contain $\eta^{''}$ in their history. Note that every other event which contains $\eta^{''}$ in its history also needs to be removed (and events containing these removed events and so on). However, since $\eta^{''}$ is the latest event in the combined histories of $\eta$ and $\eta^{'}$, no other event in their history will be removed.

Hence, $\gamma=(\F{id},o,\sigma_{\C{init}},(\Delta_r \cup \{\gamma'\}) \setminus \{\eta^{''},\eta'\} , \F{eo}_r)$ and $\gamma' = (\_, o', \sigma_{\C{init}}, \Delta_{r}^{'} \setminus \{\eta^{''}\}, \F{eo}^{'})$. Now, the length of $\tau'$ is less than or equal to $k$, and hence by the inductive hypothesis, $\gamma$ and $\gamma'$ commute modulo $\Psi$. Now, we will instantiate {\sf NON-INTERF-2} with $\eta_1 \equiv \gamma'$ and $\eta_2 \equiv \gamma$.

Specifically, $\eta_1 = (\_,o', \pi'(\sigma_{\C{init}}), \{\}, \{\}))$ where $\pi'$ is the sequence of effectors in $\Delta_r^{'} \setminus \{\eta^{''}\}$ in the order $\F{eo}^{'}$. $\eta_2 = (\_,o,\pi(\sigma_{\C{init}}), \{\eta_1\}, \{\}))$, where $\pi$ is the sequence of effectors in $\Delta_r  \setminus \{\eta^{''},\eta'\}$. Then, by the behavioral stability of $\Psi$, in the execution $C_{\C{init}} \xrightarrow{\eta_1} C_1 \xrightarrow{\eta_2} C_2$, $\eta_1$ and $\eta_2$ commute modulo $\Psi$. 

We let $\eta_3 \equiv \eta^{''}$ be the interfering event. Then, in the execution $C_{\C{init}} \xrightarrow{\eta_3} D_{1} \xrightarrow{\eta_1^{'}} D_2 \xrightarrow{\eta^{'}_{2}} D_3$ where $\eta_1^{'} = (\_,o', \pi'(\sigma_{\C{init}}, \{\eta_3\}, \{\}))$ and \linebreak
$\eta_2^{'} = (\_,o,\pi(\sigma_{\C{init}}, \{\eta_3, \eta_1^{'}\}, \{(\eta_3, \eta_1^{'})\}))$, by {\sc NON-INTERF-2}, $\eta_1^{'}$ and $\eta_2^{'}$ commute modulo $\Psi$. However, $\eta_1^{'} \equiv \eta^{'}$ and $\eta_2^{'} \equiv \eta$. Hence, by the behavioral stability of $\Psi$, $\eta^{'}$ and $\eta$ commute modulo $\Psi$. \qed

\end{proof}

\section{CRDT Definitions}
\textbf{Simple Set:}

\begin{lstlisting}
  S$\in \bbbp(E)$
Add(a):S,  $\lambda$S'.S'$\cup${a}
Remove(a):S,  $\lambda$S'.S'$\setminus${a}
Lookup(a):S,  a $\in$ S
\end{lstlisting}

\textbf{ORSet}:

\begin{lstlisting}
  S$\in \bbbp(E \times I)$
Add(a,i):S
  $\lambda$S'.S'$\cup${(a,i)}
Remove(a):S 
  $\lambda$S'.S'$\setminus${(a,i):(a,i)$\in$S}
Lookup(a):S
  $\exists$(a,i)$\in$A
\end{lstlisting}

\textbf{ORSet with Tombstones:}
\begin{lstlisting}
  S$\in \bbbp(E \times I) \times \bbbp(E \times I)$
Add(a,i,(A,R))
  $\lambda$(A',R').(A'$\cup${(a,i)},R')
	
Remove(a,(A,R)) 
  $\lambda$(A',R').(A',R'$\cup${(a,i):(a,i)$\in$A}
  
Lookup(a,(A,R))
  $\exists$(a,i)$\in$A$\wedge$(a,i)$\notin$R
\end{lstlisting}

\textbf{USet}:

\begin{lstlisting}
 S$\in \bbbp(E)$
Add(a):S 
  $\lambda$S'.if(a $\notin$ S) then S'$\cup${a} else S'
Remove(a):S
  $\lambda$S'.if(a $\in$ S and a $\in$ S') then S'$\setminus${a} else S'
Lookup(a):S
  a $\in$ S
\end{lstlisting}

\textbf{RGA}:
\begin{lstlisting}
A $\in \bbbp(E \times I \times I)$, R $\in \bbbp(I)$
AddRight(e,a,i):(A,R)
  $\lambda$(A',R'). 
    if ($\exists$ (_,e,_)$\in$A$\cup$A' and e $\notin$ R and $\forall$(_,e',_)$\in$A.e'<i) then
      (A'$\cup${(a,i,e)},R)
    else
  	  (A',R')
 
Remove(i):(A,R)
  $\lambda$(A',R'). 
    if ($\exists$(_,i,_)$\in$A) then
      (A', R'$\cup${i})
    else
      (A',R')
\end{lstlisting}

\textbf{RGA-No-Tomb}:
\begin{lstlisting}
S $\in \bbbp(E \times I \times I)$
AddRight(e,a,i):S
  $\lambda$S'. 
    if ($\exists$ (_,e,_)$\in$S$\cup$S' and $\forall$(_,e',_)$\in$S.e'<i) then
      S'$\cup${(a,i,e)}
    else
  	  S'
 
Remove(i):S
  $\lambda$S'. 
    if ($\exists$(_,i,_)$\in$S) then
      S'$\setminus${(_,i,_)}
    else
      S'
\end{lstlisting}

\textbf{2P2P-Graph}:
\begin{lstlisting}
VA $\in \bbbp(E)$, VR $\in \bbbp(E)$, EA $\in \bbbp(E \times E)$, ER $\in \bbbp(E \times E)$
AddVertex(v):(VA,VR,EA,ER)
  $\lambda$(VA',VR',EA',ER'). 
    (VA'$\cup${v},VR',EA',ER')

RemoveVertex(v):(VA,VR,EA,ER)
  $\lambda$(VA',VR',EA',ER'). 
    if(v$\in$VA$\setminus$VR and $\forall$(u,w)$\in$EA$\setminus$ER.u$\neq$v and w$\neq$v and v$\in$VA') then
      (VA',VR'$\cup${v},EA',ER')
    else
     (VA',VR',EA',ER')
      
AddEdge(u,v):(VA,VR,EA,ER)
  $\lambda$(VA',VR',EA',ER'). 
    if(u$\in$VA$\setminus$VR and v$\in$VA$\setminus$VR) then
      (VA',VR',EA'$\cup${(u,v)},ER')
    else
      (VA',VR',EA',ER')
      
RemoveEdge(u,v):(VA,VR,EA,ER)
  $\lambda$(VA',VR',EA',ER'). 
    if((u,v)$\in$EA$\setminus$ER and u$\in$VA$\setminus$VR and v$\in$VA$\setminus$VR and (u,v)$\in$EA') then	
      (VA',VR',EA',ER'$\cup${(u,v)})
    else
      (VA',VR',EA',ER')
\end{lstlisting}

\textbf{Graph-with-ORSet}
\begin{lstlisting}
V : ORSet(E), E : ORSet(E$\times$E)
AddVertex(v):(V,E)
  $\lambda$(V',E').
    (ORSet.Add(v):V(V'), E')
    
RemoveVertex(v):(V,E)
  $\lambda$(V',E').
    if(ORSet.Lookup(v):V and $\nexists$(u,w).(ORSet.Lookup((u,w)):E or 
            ORSet.Lookup((u,w)):E')  and (u=v or w=v)) then
      (ORSet.Remove(v):V(V'), E')
    else
      (V',E')
    
AddEdge(u,v):(V,E)
  $\lambda$(V',E').
    if(ORSet.Lookup(u):V and ORSet.Lookup(v):V and ORSet.Lookup(u):V' 
           and ORSet.Lookup(v):V') then
      (V',ORSet.Add((u,v)):E(E'))
    else
      (V',E')
      
RemoveEdge(u,v):(V,E)
  $\lambda$(V',E').
    (V',ORSet.Remove((u,v)):E(E'))    
\end{lstlisting}

\end{document}